
\magnification=1200

              $$ $$ $$ $$
    \centerline {\bf   Algebras with Operator and Campbell-Hausdorff
                   Formula}
                  $$ $$
       \centerline  {\bf O.M. Khudaverdian}
                  $$ $$
   {\it Department of Theoretical Physics, Yerevan State University,
    Yerevan, Armenia}\footnote {*}
      {e-mail addresse: khudian@dircom.erphy.armenia.su}

                 $$ $$
        \centerline {August 1994}
                 $$ $$

       {\it We introduce some new classes of algebras and
  estabilish in these  al\-geb\-ras Camp\-bell--Haus\-dorff
  like formula.
  We describe the application of these constructions
  to the problem of the connectivity of the Feynman graphs
  corresponding to the Green functions in Quantum Fields Theory.}
	     $$ $$ $$ $$

   \eject
  \centerline {\bf 1. Introduction}
  It is well known  Campbell-Hausdorff formula (CH formula)
 which  states that for arbitrary operators $A$ and $B$
$\log e^A e^B$ is expressed through the commutators of the operators
$A$ and $B$:
                         $$
\log e^A e^B=A+B+{1\over 2}[A,B]+{1\over 12}[A,[A,B]]+
           {1\over 12}[B,[B,A]]+\dots
                                             \eqno(1.1)
                        $$
  where  $[A,B]=AB-BA$.

 	Let $A$ be an associative algebra with unity with
   linear operator $\Pi$ acting on it such that
                    $$
       \Pi 1=1,\quad \forall a,b\in A\quad\Pi(a\Pi b)=\Pi (ab),
       \quad \Pi ((\Pi a)  b)=\Pi a \Pi b \quad.
                                 \eqno (1.2)
                     $$
  We formulate in this paper a modification of CH-formula for
 the algebras of this type and discuss one of its applications.

   For arbitrary $a_1,\dots, a_n\in A$ we define
  ${\cal G}_\Pi (a_1,\dots,a_n)$
  as the Lie algebra generated by the elements $a_1,...,a_n$
 and by the linear operator $\Pi$ in the following standard way:
  Let $[A]$ be commutators Lie algebra---
 $[A]$ is the algebra $A$ with redefined multiplication
                    $$
            [u,v]=uv-vu.
                                   \eqno (1.3)
                 $$
 Then ${\cal G}_\Pi (a_1\dots,a_n)$ is the minimal Lie subalgebra of $[A]$
      which obeys to the conditions
                 $$
  a_1,...a_n\in {\cal G},
 \quad{\rm if}\quad c\in {\cal G}\quad{\rm then}
 \quad \Pi c\in {\cal G}.
                                         \eqno (1.4)
                 $$
     $$ $$
  {\bf Theorem 1} For arbitrary $a,b\in A$
                  $$
            \log \Pi e^a e^b \in {\cal G}_\Pi(a,b)\quad .
                                                      \eqno (1.5)
                 $$
\bigskip
\indent
 {\bf Example 1}  In the case if $\Pi$ is identity operator
      ($\Pi={\bf id}$)
 then the conditions (1.2) are fulfilled automatically
     and we come to   CH formula (1.1).

       {\bf Example 2} Let $\Gamma$ be a space of the functions
   and $D$-the associative algebra of the differentiation operators
   of all the orders acting on $\Gamma$. We consider the projection
   operator $\Pi$ acting on $D$ as
                       $$
                \Pi a \colon (\Pi a)f=(a 1)f ,
                                        \eqno (1.6)
		       $$
  where $a\in D$, a function $f\in \Gamma$,
  $1$ is unity function.---$\Pi$ is the projection
 operator which extracts from the differential
 operator its null degree part. It is easy to see that
 $\Pi$ obeys to conditions (1.2).
 $$ $$
     Before going to detailed proof of this
   Theorem we briefly recall the
  algebraic proof of CH-formula (the case where $\Pi={\bf id}$)
  which is based on the following considerations
  (See for details for example [1,2]).

  Let ${\cal G}$ be an arbitrary Lie algebra
  and ${\cal U}({\cal G})$ its universal enveloping algebra.

  One can define comultiplication $\delta$ on
   ${\cal U}({\cal G})$--- the homomorphism
                        $$
   \delta\colon {\cal U}\rightarrow
    {\cal U}\otimes {\cal U}
                                                  \eqno (1.7)
		$$
 which is correctly and uniquely defined by
 its values on $\iota {\cal G}$:
		$$
   \forall x\in {\cal G}\quad  \delta \iota x=
  \iota x\otimes 1+1\otimes \iota x .
                                                    \eqno (1.8)
		$$
     ($\iota$ is canonical embedding of
 ${\cal G}$ in ${\cal U}$ (monomorphism of
 ${\cal G}$ in $[{\cal U}]$)).
 The elements $a\otimes 1+1\otimes a$ of $U\otimes U$ are called
 primitive.

 The remarkable fact is that comultiplication $\delta$ extracts
 ${\cal G}$ from $U$ ([1,2]):
                     $$
        \delta a=a\otimes 1+1\otimes a\quad {\rm iff}\quad a=\iota x.
		     						\eqno(1.9)
		$$

	In the case if $A=K(a,b)$ is free associative algebra
  with unity with two generators $a,b$ then it coincides
 with the universal enveloping algebra of the Lie algebra
 ${\cal G}(a,b)$ (${\cal G}(a,b)={\cal G}_{{\bf id}}(a,b)$
  is subalgebra of $[K(a,b)]$ defined
 by (1.4)). So we can apply (1.9) for proving (1.1)
 in the case $a,b\in K(a,b)$.

 $x\otimes 1$ and $1\otimes y$ commute in $U\otimes U$ so
 it is easy to calculate that
			$$
  \delta \log e^a e^b=\log e^{a\otimes 1+1\otimes a}
   e^{b\otimes 1+1\otimes b}=
  \log e^a e^b \otimes 1+1\otimes \log e^a e^b
  					    		 \eqno (1.10)
			$$
        is primitive in $U\otimes U$.
	(In (1.10) $x$ and $\iota x$ are identified).
  CH-formula is proved for the free algebra
  $K(a,b)$ hence for arbitrary
 associative algebra with unity.

  Using CH--formula (1.1) we can reformulate the statement of the Theorem 1:

  {\bf Theorem $1^{\prime}$} If $A$ is an associative
  algebra with unity and with linear operator
  $\Pi$ which acts on it obeying to the conditions
  (1.2) then
                   $$
       \forall c\in A\quad
     \log \Pi e^c \in {\cal G}_{\Pi}(c) \quad .
                                          \eqno (1.11)
                  $$
                  $$
     \left(
     \log \Pi e^c=\Pi c+{1\over 2}\Pi [c,\Pi c]+
     {1\over 6}\Pi [c,\Pi [c,\Pi c]]+
     {1\over 6}\Pi [\Pi c,[\Pi c,c]]+
     {1\over 6}[\Pi c,\Pi [c,\Pi c]]+
               \dots
               \right)
		$$

 (${\cal G}_\Pi(c)={\cal G}_\Pi(c,0)$ (1.5), if
  $c=\log e^a e^b\in {\cal G}_{id}(a,b)$ then
 ${\cal G}_\Pi (c)\subset {\cal G}_\Pi (a,b)$.)

  To generalize the considerations above for
  proving the Theorem $1^{\prime}$ in the Section 2
   we consider the associative algebras
  provided with additional structure
   corresponding to the action of operator
  $\Pi$ obeying to the conditions
   (1.2) (CH-algebras) and
   construct free associative CH-algebra
   with one generator.

   In the Section 3 we introduce Lie CH-algebras study their
   universal enveloping algebras. The main result of this
  paper is the Theorem 2 which is formulated and proved
 in this Section and which allows us to prove the Theorem
  $1^\prime$ in the same way like the proving of
   CH-formula ((1.7)--(1.10)).

   In the Section 4 considering  the algebras like
   in the Example 2  we use
  the Theorem 1 for the algebraic reformulation of the
 conditions of the connectivity of Feynman diagrams corresponding to
 the logarithm of partition function in quantum field theory. It
 was the considerations which stimulated us for this algebraic
 investigation. (The example 2 was the basic example for formulating
 the conditions (1.2)).

   The CH--algebras which are introduced in this paper
 seems to be interesting in the applications.
 Professor V.M.Buchstaber offers the general construction
 for CH--algebras. He noted also that so called Novikov's
 O--doubles [3] (which are the natural generalization of the algebra
 constructed in the Example 2) are the interesting examples
 of CH--algebras. We consider these examples in the Section 5.

  We have to note that in the formulae (1.1), (1.5), (1.11)
 the expressions $\log e^a e^b$, $\log \Pi e^a e^b$, $\log \Pi e^c$
  are considered as formal power series corresponding to the functions
 $\log$, $e$. All the statements have the sense for arbitrary large
  but finite initial sequences of these formal series.

   All the algebras considered here are the linear spaces on the
  real (or complex numbers). The associative
 algebras considered here are the associative algebras with unity.

 $$ $$ $$ $$
 \centerline {\bf 2. CH--algebras}
  We call the pair $(A,\Pi)$---associative CH--algebra
 (with unity) if $A$ is the associative algebra (with unity) and
$\Pi$ is linear operator on it obeying to the conditions (1.2).
We say that operator $\Pi$ provides the algebra $A$ with CH-structure.
  From (1.2) it is evident that $\Pi$ is projection operator
  $(\Pi^2=\Pi)$ and ${\bf Im}\Pi$ is subalgebra in $A$.

  For example the linear operator $\Pi$(1.6) on the algebra $D$ defined
 in the example 2 provides this algebra by CH-structure.

 Of course every associative algebra can be provided with
trivial CH-structure ($\Pi={\rm {\bf id}} $).

 The homomorphism $\varphi$ of the associative algebra $A_1$ in
the associative algebra $A_2$ is the morphism of corresponding
CH-algebras (CH-morphism)
 $\varphi: (A_1,\Pi_1)\rightarrow (A_2,\Pi_2)$ if
                      $$
             \varphi \circ\Pi_1=\Pi_2\circ\varphi \quad.
                                                             \eqno (2.1)
                      $$
 We need also the construction of tensor product of CH-algebras:
 $(A_1,\Pi_1)\otimes (A_2,\Pi_2)=
 (A_1\otimes A_2, \Pi_1\otimes \Pi_2)$ where
                    $$
       (\Pi_1\otimes \Pi_2)\,(a_1\otimes a_2)=
      \Pi_1 a_1\otimes \Pi_2 a_2 .
                                                     \eqno (2.2)
                     $$

 The CH-algebra $(A_\pi (L),P)$ is free algebra with one generator
in the category of
CH--associative algebras with unity if for arbitrary CH--algebra
$(B,\Pi)$ from this category and for arbitrary $c\in B$ there
exists unique CH-morphism
 $\varphi: \quad(A_\pi(L),P)\rightarrow (B,\Pi)$ such that
                  $$
            \varphi (L)=c\, .
                                           \eqno (2.3)
                  $$

{\bf Proposition 1}  There exists unique (up to isomorphism) free
CH-associative with unity algebra with one generator.

   We give briefly the construction of this algebra.

 Let $\pi,L$ be two formal symbols.

 ${\cal A}_0$ is the alphabet containing one letter $L$.
$\Gamma_0$ is the semigroup of the words on the alphabet ${\cal A}_0$.
(To the empty word corresponds the unity in $\Gamma_0$.)
 Let ${\cal A}_1$ is the alphabet containing the letter
 $L$ and all the letters $\pi s$ where $s\in \Gamma_0$ and $s\neq 1$.
 $\Gamma_1$ is the semigroup
 of the words on the alphabet ${\cal A}_1$. By induction
the alphabet ${\cal A}_{n+1}$ contains all the letters of the alphabet
${\cal A}_n$ and  the new letters $\pi s$ where
$s\in \Gamma_n\setminus\Gamma_{n-1}$, $\Gamma_{n+1}$ are the words on the
${\cal A}_{n+1}$. The semigroup
                       $$
    \Gamma=
 \Gamma_0\cup\Gamma_1\cup\dots\cup\Gamma_n\cup\dots
                                                       \eqno (2.4)
                       $$
 is the semigroup of the words on the alphabet
                      $$
 {\cal A}={\cal A}_0\cup\dots\cup{\cal A}_n\cup\dots
                                                         \eqno (2.5)
                      $$
 The linear combinations of the words of the semigroup
  $\Gamma$ with the coefficients from the field of
real (or complex) numbers consist
 the associative with unity algebra $A(\pi,L)$.
 On this algebra one can
 consider the linear operator $P$ which is
 defined on $A(\pi,L)$ by its action on the words from $\Gamma$:
                $$
           Pw=\pi w.
                                                           \eqno (2.6)
                $$
  If $w$ is the word from $\Gamma_k$ then $\pi w$ is the letter
 from ${\cal A}_{k+1}$(the oneletter word from $\Gamma_{k+1}$).

 Now we construct the  algebra $A_\pi(L)$ as factor algebra of
 $A(\pi,L) $.

 Let us consider the set of ideals in the  $A(\pi,L)$:

  $J_0$ is the two--sided ideal
 ge\-ne\-ra\-ted by all the e\-le\-ments
  $P(aPb)-P(ab)$, $P(Pab)-PaPb$
  ($\forall a,b \in A(\pi,L)$) of the algebra
 $A(\pi,L)$. $J_1$ is the two--sided
 ideal ge\-ne\-ra\-ted by all the e\-le\-ments
 $Pa_0$ ($\forall  a_0\in J_0$).
  By induction $J_{n+1}$ is the
 two--sided ideal generated by all the elements $Pa_n$
  ($\forall a_n\in J_{n}$). Then
              $$
     A_\pi(L)=A(\pi,L)/ J
                                             \eqno (2.7)
              $$
where
              $$
   J=J_0\oplus J_1\oplus\dots\oplus J_n\oplus\dots
                                                     \eqno(2.8)
                 $$
   It is easy to see that  the operator $P$ defined
 on the algebra $A(\pi,L)$ by (2.6)
 is correctly defined on the algebra
 $A_\pi(L)$, provides this algebra
  with CH-structure and CH-algebra $(A_\pi(L),P)$ is
  free CH-algebra with one generator.

  We can also describe the basis of the algebra $A_\pi(L)$ (considering it
 as a linear space). We call the letters $L$,
 $\pi L$ and all the
 letters of the type $\pi(LwL)$ the regular letters
 (where $w$
 is some word in $\Gamma$).
  We call the word in $\Gamma$ regular if it contains
only regular letters. It is easy to see (using the conditions
(1.2)) that  every element of $A(\pi,L)$
 is equivalent in $A_\pi(L)$ to the
linear combination of the regular words.
  Moreover one can show that
{\it the regular words consist the linear basis of the $A_\pi(L)$}.
 We do not give here the proof of this statement. For our purposes
 it is enough only that $A_\pi(L)$ is not trivial.

 $$ $$
 \centerline {\bf  3. Lie CH-algebras} Now we introduce
  Lie CH-algebras.
 We say that the pair $({\cal G},M)$ is Lie CH-algebra if ${\cal G}$
is  Lie algebra and $M$ is linear operator on it  such that
                $$
      M^2=M,\quad \forall a,b\in {\cal G}\quad
       M[a,b]=M[a,Mb]+M[Ma,b]-[Ma,Mb]\quad.
                                                     \eqno(3.1)
                $$
   From (3.1) it follows that ${\bf Im}M$:
                    $$
        \tilde{{\cal G}}=\{{\cal G}\ni a\colon a=Ma\}
                                                      \eqno (3.2)
                    $$
    is subalgebra in ${\cal G}$.

    In the same way like for associative algebras every Lie algebra
   can be provided with trivial CH-structure ($M={\bf {\rm id}}$).

    It is easy to see that
   if $(A,\Pi)$ is associative CH-algebra
  then the operator $\Pi$ on $[A]$
  obeys to (3.1) hence $([A],\Pi)$ is Lie CH-algebra
  ($[A]$ is commutators Lie algebra (1.3)).

  In the same way as for associative CH-algebras
 the homomorphism $\varphi$ of the Lie algebra ${\cal G}_1$ in
the Lie algebra ${\cal G}_2$ is the morphism of corresponding
 Lie CH-algebras (CH-morphism)
 $\varphi: ({\cal G}_1,M_1)\rightarrow ({\cal G}_2,M_2)$
 if
                      $$
             \varphi \circ M_1=M_2\circ\varphi .
                                                             \eqno (3.3)
                      $$

   Now we ge\-ne\-ra\-lize the con\-struc\-tion of the
 u\-ni\-ver\-sal en\-ve\-lo\-ping al\-geb\-ra for CH--algebras.

     Let $({\cal U},\Pi)$ be associative CH-algebra with unity
     and $\iota$:
     $({\cal G},M)\rightarrow ([{\cal U}],\Pi)$
     is CH-morphism.

    We say that $({\cal U},\Pi)$ with CH-morphism $\iota$ is universal
    enveloping CH-algebra of Lie CH-algebra $({\cal G},M)$
    if for arbitrary associative
  CH-algebra with unity $(A,S)$
    and for the arbitrary CH-morphism $\varphi$ of $({\cal G},M)$
     in $([A],S)$ there exists unique (enveloping) CH-morphism
     $\rho$ of $({\cal U},\Pi)$ in $([A],S)$ such that
                         $$
              \rho \circ\iota=\varphi\, .
                                                          \eqno (3.4)
                        $$
  $$ $$
     If $({\cal G}_P(L),P)$ is  CH-Lie algebra generated in the
  $([A_\pi(L)],P)$ (See  Proposition 1)
  by the element $L$ and linear operator $P$ (1.4)
 then  it is easy to see that

    {\bf Proposition 2} The free CH-algebra $(A_\pi (L),P)$ is universal
 enveloping algebra for Lie CH-algebra $({\cal G}_P (L),P)$ ,
 (morphism $\iota$ is canonical embedding.)

  (The enveloping homomorphism $\rho$ is defined by the condition
 $\rho(L)=\varphi (L)$.)

  {\bf Theorem 2} For Lie CH-algebra $({\cal G},M)$ there exists unique
 (up to isomorphisms)
 universal enveloping CH-algebra $({\cal U},\Pi)$ with CH--morphism
 $\iota \colon ({\cal G},M)\rightarrow ([{\cal U}],\Pi)$
  where ${\cal U}$ with $\iota$ is universal enveloping algebra for
 Lie algebra ${\cal G}$ in usual sense.

 {\bf Corollary 1}. Let ${\cal U}$ be universal enveloping
 algebra of the Lie algebra ${\cal G}$ and
 $\rho$ be enveloping homomorphism of the
 homomorphism $\varphi\colon {\cal G}\rightarrow  [A]$
 (for some associative algebra $A$).
   If ${\cal G}$ and $A$ can be provided with CH-algebras
 structures in a way that $\varphi$ becomes CH-morphism,
 $\varphi\colon ({\cal G},M)\rightarrow  ([A],S)$
 then $\rho$ is CH-morphism too:
 $\rho\colon ({\cal U},\Pi)\rightarrow (A,S)$.

  {\bf Corollary 2} If $({\cal U},\Pi)$ is universal
 enveloping algebra of
 CH-Lie algebra $({\cal G},M)$ then comultiplication $\delta$
defined on ${\cal U}$ by (1.8) commutes with action of operator
$\Pi$:
                      $$
             \delta\Pi=(\Pi\otimes\Pi)\delta \quad.
                                                    \eqno(3.5)
                     $$

    Indeed comultiplication $\delta$ is enveloping homomorphism
 of the homomorphism
 $\varphi$ of ${\cal G}$ in $[{\cal U}\otimes {\cal U}]$
  $\colon \varphi(x)=\iota x\otimes 1+1\otimes \iota x$.
  It is easy to see that $\varphi$ is
 morphism of Lie CH-algebra $({\cal G},M)$ in Lie CH-algebra
 $([{\cal U}\otimes{\cal U}],\Pi\otimes\Pi)$. It follows from Corollary 1
 that $\delta$ is CH-morphism of $({\cal U},\Pi)$ in
 $([{\cal U}\otimes{\cal U}],\Pi\otimes\Pi)$ hence (3.5) is satisfied.

   Now  using Theorem and Corollary 2 we can
prove Theorem $1^\prime$ using (1.9).
 Indeed from Proposition 2 and Theorem 2 it
 follows that $A_\pi (L)$
 {\it is universal enveloping algebra for Lie algebra} ${\cal G}_P(L)$
 {\it (in usual sense)}.  Using Corollary 2 we can easy
check (like in (1.10)
 that
$\delta \log Pe^L$ is primitive in ${\cal U}\otimes{\cal U}$.
  Theorem $1^\prime$ is proved for free CH-algebra, so for arbitrary
 CH-algebra .

 Indeed it is easy to see that we
proved  little more:
                $$
         \log Pe^L \in
  \tilde {\cal G}_P(L)={\bf Im} P
                                                 \eqno(3.6)
                 $$

  Now we prove Theorem 2.

   {\bf Lemma}. If ${\cal U}$ is the universal enveloping algebra
 of the Lie algebra ${\cal G}$ and the linear operator $M$
 provides  Lie algebra ${\cal G}$ with CH--algebra
 structure then there exists
 the linear operator $\Pi$ providing ${\cal U}$ with CH-algebra structure
  in a way that  canonical embedding $\iota$
  ($\iota\colon {\cal G}\rightarrow [{\cal U}]$) becomes
  CH-morphism:

                      $$
                \Pi\circ\iota=\iota\circ M\, .
                                                        \eqno (3.7)
                      $$

  {\bf Remark} If Lie algebra ${\cal G}$ is provided with trivial
 CH--structure $M \colon {\bf Im}M=0$ then $\Pi$ on
 ${\cal U}$ is corresponded to augmentation
 $\varepsilon \colon {\cal U}\rightarrow {\it k}$ (counity)
 ($\Pi a = \varepsilon (a)\cdot 1$,
 $(\Pi\otimes {\bf id})\delta a=1\otimes a$ where $\delta$ is
 comultiplication (1.7)) on the Hopf algebra ${\cal U}$.

   We prove Lemma later.

  Now we prove that CH-algebra $({\cal U},\Pi)$ obeying to the conditions
 of Lemma is the universal enveloping algebra for $({\cal G},M)$.

  Let $(A,S)$ be associative CH-algebra and $\varphi$ CH-morphism of
  $({\cal G},M)$ in $([A],S)$. Because
   ${\cal U}$ is universal enveloping algebra of
   ${\cal G}$ (in usual sense)
   then there exists
  unique enveloping homomorphism  $\rho$
  of ${\cal U}$ in $A$ ($\rho\iota=\varphi$).
 We have to prove that
  $\rho$ is CH-morphism of $({\cal U},\Pi)$ in $(A,S)$:
                   $$
                \rho\Pi=S\rho.
                                                       \eqno (3.8)
                   $$
   ( The uniqueness
 of the $({\cal U},\Pi)$ is provided by the fact that if
   $(\tilde{{\cal U}},\tilde{{\Pi}})$ is another universal enveloping
 algebra of the CH-Lie algebra $({{\cal G}},M)$ then considering
 the CH-morphisms
 $\tilde{\rho}\colon(\tilde{{\cal U}},\tilde{{\Pi}})\rightarrow
   ({{\cal U}},{{\Pi}})$ and
$\rho\colon {{\cal U}},{{\Pi}}) \rightarrow
 (\tilde{{\cal U}},\tilde{{\Pi}})$ which envelop correspondingly
 the embeddings
$\iota\colon ({\cal G},M) \rightarrow
 ([{\cal U}],{{\Pi}})$ and
$\tilde {\iota}\colon ({{\cal G}},M) \rightarrow
 ([\tilde{{\cal U}}],\tilde{{\Pi}})$
 we see that $\rho\tilde{\rho}={\bf id}$,
  $\tilde{\rho}\rho={\bf id}$.)

       We prove (3.8) by induction. We consider
       the linear subspaces ${\cal U}_n$ in ${\cal U}$:
       ${\cal U}_0=1$, ${\cal U}_1=\iota {\cal G}$, ${\cal U}_n$
       ($n\geq 2$) contains all the elements
       which can be represented as linear combinations of the products
       of less than $n+1$ elements of ${\cal U}_1$.

  Inductive suggestion.---For $n\leq k$ ($n\geq 1$)
                    $$
      \forall a\in {\cal U}_n, \Pi a=\sum \iota M x_i\cdot  b_i,\quad
      {\rm where}\quad b_i\in {\cal U}_{n-1},
           x_i \in {\cal G}
                                                   \eqno (3.8a)
                   $$
             \centerline {and}
                  $$
     \forall a\in {\cal U}_n ,\quad\rho (\Pi a)=S\rho (a).
                                                    \eqno (3.8b)
                 $$
    Here and in the following we use that in a
  universal enveloping algebra
  $\iota x\cdot \iota y - \iota y\cdot  \iota x=\iota [x,y]$.
 For $k=1$ (3.8a,3.8b) are evident: $\Pi a=\Pi \iota x=\iota M x$
 by (3.7) and
 $\rho(\Pi a)=\rho\Pi (\iota x)=\rho (\iota M x)=
 \varphi (Mx)=S\varphi (x) = S\rho (\iota x)=S\rho(a)$ .

  For proving (3.8a) in the case $k\rightarrow {k+1}$ we note using (1.2)
  that
  if ${\cal U}_{k+1}\ni a=a^\prime\cdot  \iota x$
  (where $a^\prime\in {\cal U}_k$)
  then  $\Pi a=\Pi (a^\prime\cdot  \iota x)=$
  $\Pi(a^\prime\cdot  \Pi \iota x)=
  \Pi(a^\prime\cdot  \iota Mx)=\Pi(\iota Mx\cdot  a^\prime+
  a^{\prime \prime})=
  \iota M x\cdot  \Pi a^\prime+\Pi a^{\prime \prime}$
  (where $a^{\prime\prime}\in {\cal U}_k$). (3.8a) is proved.

    Using (3.8a) we prove (3.8b) for $k\rightarrow {k+1}$. Again using
    (1.2) we have that if ${\cal U}_{k+1}\ni a=\iota x\cdot  b$ then
    $\rho \Pi a=\rho\Pi(\iota x\cdot  b)=\rho\Pi(\iota x\cdot   \Pi b)=$
    (by 3.8a) $=\sum\rho\Pi(\iota x\cdot \iota My_i\cdot  c_i)=$
    (where $c_i\in{\cal U}_{k-1}$)
    $=\sum \rho\Pi(\iota[x,My_i]\cdot  c_i)+\sum\rho\Pi(\iota M y_i\cdot \iota
x
    \cdot  c_i)$=
    $\sum \rho\Pi(\iota[x,M y_i]\cdot  c_i)+
    \sum \rho\iota My_i\cdot \rho\Pi(\iota x\cdot  c_i)=$
    (by inductive suggestion (3.8b))
    $=\sum S\rho(\iota[x,M y_i]\cdot  c_i)+
     \sum S\rho\iota y_i\cdot  S\rho(\iota x\cdot  c_i)=
      \sum S\rho(\iota[x,M y_i]\cdot  c_i)+
       \sum S\rho(\iota My_i\cdot  x\cdot  c_i)=
       S\rho(\iota x\cdot  b)=S\rho (a)$.

     The Theorem 2 is proved.

     Now we prove the  Lemma.

     Let ${\cal G}$ be Lie algebra and the linear operator $M$
     provides it by CH--algebra structure (3.1).

    Let $\{b_\alpha,e_i\}$
    ($\alpha \in I_0,i\in I_1$) be basis in ${\cal G}$ such that
    $\{b_\alpha\}$ is basis in subalgebra
   $\tilde{{\cal G}}={\bf Im} M$
   (3.2)  ($e_i\not\in\tilde{{\cal G}}$).
     We assume that $I=(I_0,I_1)$ is well-ordered
  and the elements of $I_0$
     precede those of $I_1$.
      The monomials
      $\{b_{\alpha_1\dots\alpha_n}\cdot  e_{i_1\dots i_m}\}$
      where $\alpha_1\preceq\dots\preceq\alpha_n,
            i_1\preceq\dots\preceq i_m$ and
      $b_{\alpha_1\dots\alpha_n}=
      \iota b_{\alpha_1}\cdot \dots\cdot  \iota b_{\alpha_n}$,
      $e_{i_1\dots i_m}=
      \iota e_{i_1}\cdot \dots \cdot \iota e_{i_m}$
      is the basis (Birchof de Witt basis ) of ${\cal U}$.

     We consider the filtration on  ${\cal U}$:
                          $$
    {\cal U}_{(0)}\subset{\cal U}_{(1)}\subset\dots\subset
             {\cal U}_{(n)}\subset\dots
                                                        \eqno(3.9)
                          $$
       where ${\cal U}_{(k)}$ is  the linear combination of the  basis
        elements
        $\{b_{\alpha_1\dots\alpha_n}\cdot  e_{i_1\dots i_m}\}$ for $m\leq k$.

    (${\cal U}_{(0)}$ is the universal enveloping algebra
   of $\tilde {{\cal G}}$ defined by (3.2).)

  We note using (3.1) that Lie algebra ${\cal G}$ is
 the sum (as the linear space) of two subalgebras:
		       $$
     {\cal G}=\tilde {{\cal G}} \oplus \tilde {\tilde{{\cal G}}},\quad
     (\tilde {{\cal G}}={\bf Im} M,\,
     \tilde {\tilde{{\cal G}}}={\bf ker} M,\,
     \tilde {{\cal G}}\cap \tilde {\tilde{{\cal G}}}=0)\, .
					   \eqno (3.10)
		       $$
    If $\{e_i\}$ is a basis in the Lie subalgebra
     $\tilde {\tilde{{\cal G}}}$:
		       $$
	       	     Me_i=0
						 \eqno(3.11)
		       $$
  then it is easy to see that the linear operator $\Pi$:
		       $$
     \Pi(b_{\alpha_1\dots\alpha_n}\cdot  e_{i_1\dots i_m})=0,\,
     {\rm if}\, m\geq 1,\,
        \Pi\Big \vert_{{\cal U}_{(0)}}={\bf id},\,
        (\Pi\colon {\cal U}\rightarrow{\cal U}_{(0)})\,
						\eqno(3.12)
		      $$
  is satisfied to the conditions of Lemma.

  (In the case if the condition (3.11) does not hold one can show
 (by induction) that operator $\Pi$ can be defined as
 $\Pi=\Phi^{\infty}$ where
 $\Phi(b_{\alpha_1\dots\alpha_n}\cdot  e_{i_1\dots i_m})=$
 $b_{\alpha_1\dots\alpha_n}\cdot  e_{i_1\dots i_{m-1}}
 \cdot  \iota Me_{i_m}$,
 $\Phi\Big\vert_{{\cal U}_{(0)}}={\bf id }$,
 ($\Phi\colon{\cal U}_{(n+1)}\rightarrow{\cal U}_{(n)}\,
      {\rm if}\, n\geq 1$,
  if $a\in{\cal U}_{(k)}$  then
     $\forall N\geq k$
       $\Phi^{N+1}a=\Phi^N a\in {\cal U}_{(0)}$.)

  The proof is finished.

   {\bf Note} Our aime was to give a pure algebraic proof of the
  formula (3.6). However we want to note that
  using the above algebraic statements (Proposition 2, Theorem 2)
 and the formulae (3.10--3.12) one can give
 another proof of (3.6) which is based on the following
 fact yielded from CH--formula (1.1):
  $e^x$ is decomposed into the product
  $e^{\tilde x}e^{\tilde {\tilde x}}$ where
  ${\tilde x}\in {\bf Im}M$, ${\tilde {\tilde x}}\in {\bf ker}M$.

        $$ $$
      \centerline   {{\bf  4.Application}}
     The statement of the Theorem 1 can be used for
 investigation the problem of the connectivity of the Feynman
 graphs corresponding to the Green functions in Quantum Fields
 Theory.

 In Quantum Field Theory  it is well known the
 Theorem about the connectivity of the Feynman graphs corresponding
 to the logarithm of partition function (PFLC Theorem)---
 generating functional of the Green functions.
 (See for details for example [4].)

 The Green functions  of the quantum theory are the vacuum
 expectation values of the time ordered products
 of field operators:
                     $$
     G(x_1,\dots,x_n)=
  <T(\hat {\phi}(x_1) \dots \hat {\phi}(x_n))>,
                                                         \eqno(4.1)
                      $$
 where the classical theory is defined by the classical action
 $S(\phi)$---the functional on the classical fields $\phi(x)$
 corresponding to the fields operators $\hat{\phi}(x)$.

 The Green functions can be collected together in the generating
 functional $Z(J)$ (partition function)
 --- a formal power series on the "classical
 sources" $J(x)$:
                    $$
     Z(J)=\sum_{N=0}^{\infty}{1\over N!}
     \int G(x_1,\dots,x_N)J(x_1)\dots J(x_N)dx_1\dots dx_N.
                                                         \eqno (4.2)
                    $$
   In the case where $S(\phi)$ is the action of free theory
                         $$
     S(\phi)= S(\phi)_{\rm free}=\int \left(
         {1\over 2}\phi (x)K(\partial)\phi(x)
               \right)dx
                                                           \eqno(4.3)
                   $$
 where $K(\partial)$ is some invertible differential operator
 (for example $K=\partial^2$) the functional $Z(J)$ can be easily
 calculated:
                                $$
  Z(J)_{\rm free}=e^{\int J(x)\Delta(x-y)J(y)dxdy}\, ,
                                                        \eqno (4.4)
                           $$
 where $\Delta(x-y)$ is two-point Green function of the free theory
 which is obtained by inverting operator $K(\partial)$
                       $$
       K(\partial)\Delta(x,y)=\delta(x-y) \, .
                        $$
    In the case of full interacted theory where
                     $$
                   S(\phi)=
               S(\phi)_{\rm free}+
                S(\phi)_{\rm int}
                     $$
  the functional $Z(J)$ is given by the following formal
  expression
                            $$
  Z(J)=e^{  S_{\rm int}({\delta\over \delta J})}
                Z_{\rm free}=
 e^{  S_{\rm int}({\delta\over \delta J})}
        e^{\int J(x)\Delta(x-y)J(y)dxdy}.
                                                 \eqno(4.5)
                           $$
   It is well known Gell-Mann and Low formula which leads to
 the perturbative expansion of the Greens functions in terms of
 Feynman graphs [4]. To every monom in a power
 series expansion of (4.5) by $J$
 correspond Feynman graphs connected or disconnected.

 {\bf PFLC--Theorem}. In the functional $\log Z(J)$
 give contribution only connected Feynman graphs.

  (The analogous statement is in Statistical Physics where
 to $\log Z$ correspond the free energy of the system and in the
 Probability Theory where to $\log Z(J)$ correspond semyinvariant [5]).

 The standard proofs of this Theorem are based on the  recursive
 procedure of the Feynman graphs investigation.

  We discuss PFLC--Theorem using the Theorem 1.

   We rewrite (4.5) symbolically
                   $$
          Z=e^{\hat{T}}e^K.
                                                        \eqno (4.6)
                     $$
 To consider $\hat T$ and $K$ on an equal footing we rewrite
 (4.6):
                        $$
           Z=\Pi e^{\hat{T}}e^{\hat{K}}
                                              \eqno(4.7)
                      $$
       where $\Pi {\hat{A}}=\hat {A}1$ and $\hat K$
 is the operator of the multiplication on $K$
 (compare with Example 2 of the Section 1).

 To every  element $a$ of the associative
 CH-algebra $(A_{\Pi}(\hat T,\hat K),\Pi)$ generated by
 $\hat{T}$, $\hat K$ and projection operator $\Pi$
 correspond Feynman graphs connected or disconnected.
 (For example to the element $c=\Pi a \Pi b$
 correspond disconnected Feynman graphs if
 $\Pi a$ and $\Pi b$ are not trivial elements
 of the $A_{\Pi}(\hat T,\hat K)$.)
  From the Theorem 1 it follows that the PFLC--Theorem
 is reduced to the algebraic statement:
 {\it to the elements of the
 Lie CH-algebra $({\cal G}_\Pi (\hat T,\hat K),\Pi)$
 correspond the connected Feynman graphs.}
  This statement is right
  as far as to $\hat T$ and $\hat K$
 correspond connected Feynman graphs (which takes place for field
 theory standard lagrangians). One can show it noting
 that to the commutator operation in the
 $({\cal G}_\Pi (\hat T,\hat K)$
 corresponds the gluing of the corresponding
 Feynman subgraphs.

      $$ $$
  \centerline {\bf {5. Discussions}}

  To find algebraic reformulation
 of the PFLC-Theorem we considered associative
 algebras provided with additional operation
 by means of linear operator $\Pi$ which obeys to the conditions (1.2)
 and the Lie algebras corresponding to them (CH-algebras).
  We show that these algebras have the properties similar
 to usual ones (Theorem 2). In particular one can formulate
 Campbell-Hausdorff like statement (Theorem 1).

  It is interesting to study nontrivial examples of CH--algebras.
  Following to V.M. Buchstaber we consider

 {\bf Example 3} [6]. Let $A$ be associative algebra whith unity
 which is the left $B$--module ($B$ is subalgebra in $A$) and $A$
 admits the expansion
 $A=B+\sum_{k\geq 1} B\cdot  a_k$
 in a way that
 $a_k a_l =\sum_{q\geq 1}b_{kl}^q a_q$ where
 $b_{kl}^q\in B$.
 Then it is easy to see that
 $\Pi\colon A\rightarrow B\colon a=\sum_{k\geq 0}b_k a_k\rightarrow  b_0$
 ($a_0=1$) provides $A$ with CH-structure.
  (Compare with 3.9--3.12.)

  One can show that every associative CH-algebra
  $(A,\Pi)$ can be represented
 in this way using that
  $C={\bf ker}\Pi$ is the subalgebra which is as well as
  $A$ the left module over the subalgebra
  $B={\bf Im}\Pi$ (as it follows from (1.2)).

  The interesting example of this construction is

 {\bf Example 4} [6]. Let $X$ be the Hopf algebra with
 comultiplication $\delta$ and augmentation
 $\varepsilon\colon X\rightarrow{\it k}$.
  (Compare with {\bf Remark} after {\bf Lemma}.) Let
 an algebra ${\cal M}$ be left $X$--module such that
  $\forall x\in X$ and $\forall u,v\in M$
  $x(uv)=\sum_i x^1_i(u)\cdot x^2_i(v)$ if
 $\delta x=\sum_i x^1_i\otimes  x^2_i$. (Milnor module.)
 The algebra $A={\cal M}X$ of the linear combinations
 $(A\ni a=\sum u_k x_k,\quad u_k\in {\cal M},x_k\in X)$
 is so called Novicov's {\bf O--double} [3].
 (The multiplication is defined by
 $(ux)\cdot(vy)=u\sum_i x^1_i(v)x^2_i y$ where
 $\delta x=\sum_i x^1_i\otimes x^2_i$.)
 If $\{e_\alpha\}$ is basis in $X$ such that
 $\varepsilon e_0=1$ and
 $\varepsilon  e_\alpha=0$ for $\alpha\neq 0$ then
 the linear operator $\Pi$ defined on O--double
 ${\cal M}X$ by the condition
 $\Pi(\sum_{\alpha} u_\alpha e_\alpha)=u_0 e_0$
 provides ${\cal M}X$ with natural CH--structure.

  The CH--algebra $({\cal M}X,\Pi)$ is on one hand
 the natural generalization of the algebra described
 in Example 2 where $X$ is the algebra of differentiations
 with constant coefficients and ${\cal M}$ is the algebra
 of the functions. On other hand this algebra arises
 in the different applications.
 The model example for O--double is the algebra
 $A^U=\Lambda X$
 of cohomological operations in complex cobordism theory [7]
 where $X$ is so called "Landweber-Novikov" algebra [7,8],
 $\Lambda$ is the $U$--cobordism ring "for the point".
 The algebra $A^U$ is related to the
 differential operators on some infinite--dimensional
 Lie group.---It was shown in [9] that if
  ${\it Diff}({\bf R})$ is the group of the
 diffeomorphisms of the real line ${\bf R}$ and
  $G$ its subgroup:
  $G=\{{\it Diff}\ni f\colon f(0)=0,\,f^{\prime}(0)=1\}$
  then
 $X$ is isomorphic to the universal enveloping algebra
 ${\cal U}({\cal G}(G))$ of the Lie algebra
 ${\cal G}$ of the group $G$. $\Lambda$ is the ring
 of polinomials on this group. We see that
  $A^U$ returns us again to the example 2.

  It is interesting to study in details these
  and other examples of CH--algebras.

 On other hand it is interesting to study how much  the statements
 of the Theorems 1 and 2 depend on the conditions (1.2) on the linear
 operator $\Pi$.

   The considerations of the 4-th section lead us in fact
 to redefinition of the connectivity conditions as the
 conditions of belonging to specially constructed Lie algebra.
  It can be interesting to analyze in details
 the relations between this definition and usual ones.

  $$ $$

              \centerline{\bf Acknowledgement}

   I am deeply grateful to S.G.Dalalian who encouraged and stimulated
   the algebraic investigations in Sections 2 and 3.

  I am also grateful to Grigorian S. and Nahapetian B. for useful
 discussions.

  During the highly fruitful discussions of the
 basic results of this paper professor V.M.Buchstaber
 offered me the important examples of CH--algebras.
 (The examples 3 and 4 in 5--th section). I am deeply
 grateful him for it.

  This work was partially supported by
 International Science Foundation (Soros found): grant M3Z000.
      $$ $$
     \centerline {\bf References}

  1. Bourbaki N.  Elements de Mathematique.-- Groupes et
 Algebres de Lie. Chapitre 1,2.-- Paris "Herman", 1971,1972.

  2. Postnikov M.M.  Lectures on Geometry.--Semestre 5,
    Lie Groups and Al\-geb\-ras.-- Mos\-cow,"Na\-u\-ka",
   pp. 99-121, (1982) (in Russian).

  3 Novikov S.P. Different doubles of the Hopf Algebras.
  Operator algebras on the Quantum Groups,
  Complex Cobordisms.--- Uspechi Math Nauk, {\bf 47},iss 5,
  p.189-190 (1992) (in Russian).

  4  C. Itzykson, J.-B. Zuber   Quantum Field Theory.
         McGraw-Hills, 1980.

  5   V.A. Malishev, R.A. Minlos.-- Gibbs random fields.
    "Nauka", Moscow, 1985

  6  V.M. Buchstaber.  Private communication.

  7 Novikov S.P.--- Isv. AN SSSR (ser math) {\bf 31} pp.855-951 (1967)
	  (in Russian)

  8 Landweber P.--- Trans. Amer.Math. Soc.,{\bf 27},
  n1, pp.94-110 (1967)

  9 Buchstaber V.M., Shokurov V.S.---  Funk.  Analiz i priloz.
  {\bf 12}, iss 3, pp.1-14 (1978) (in Russian)

 \bye